\documentclass{ws-procs10x7}
\usepackage{balance}

\newcolumntype{d}[1]{D{.}{.}{#1}}

\def\Journal#1#2#3#4{{\it #1} {\bf #2}, #3 (#4)}

\makeindex
\begin{document}

\title{RELATIVISTIC DESCRIPTION OF SEMILEPTONIC DECAYS\\ OF HEAVY BARYONS}

\author{D. Ebert}

\address{Institut f\"ur Physik, Humboldt--Universit\"at zu Berlin,
Newtonstr. 15, D-12489  Berlin, Germany}

\author{R. N. Faustov and V. O. Galkin}

\address{Dorodnicyn Computing Centre, Russian Academy of Sciences,
  Vavilov Str. 40, 119991 Moscow, Russia}


\twocolumn[\maketitle\abstract{Semileptonic decays of heavy baryons
  consisting of one heavy 
($Q=b,c$) and two light ($q=u,d,s$) quarks are considered in the
heavy-quark--light-diquark approximation. The relativistic
quasipotential equation is used for obtaining masses and wave
functions of both diquarks and baryons within the constituent
quark model. The weak transition matrix elements are expressed
through the overlap integrals of the baryon wave functions. 
The Isgur-Wise functions are determined in the whole accessible
kinematic range. The exclusive semileptonic decay rates are calculated
with applying the heavy quark $1/m_Q$ 
expansion. The evaluated $\Lambda_b\to \Lambda_c l \nu$ decay 
rate agrees with its experimental value.}
]

We study the semileptonic decays of heavy baryons in the
heavy-quark--light-diquark approximation using relativistic bound
state equation and heavy quark expansion \cite{efghbm,efg}. Baryons containing both
scalar and axial vector diquarks are considered.  In order to
calculate the heavy baryon decay rate it is necessary to determine the
corresponding matrix element of the  weak current between baryon states:   
\begin{eqnarray}\label{mxet} 
&&\!\!\langle B_{Q'}(p_{B_{Q'}}) \vert J^W_\mu \vert B_Q(p_{B_Q})\rangle
=\\
&&\!\!\int \frac{d^3p\, d^3q}{(2\pi )^6} \bar \Psi_{B_{Q'}\,{\bf p}_{B_{Q'}}}({\bf
p})\Gamma _\mu ({\bf p},{\bf q})\Psi_{B_Q\,{\bf p}_{B_Q}}({\bf q}),\nonumber
\end{eqnarray}
where $\Gamma _\mu ({\bf p},{\bf
q})$ is the two-particle vertex function and  
$\Psi_{B\,{\bf p}_B}$ are the
baryon ($B=B_Q,B_{Q'})$ wave functions projected onto the positive
energy states of 
quarks and boosted to the moving reference frame with momentum ${\bf p}_B$.

The hadronic matrix elements for the semileptonic decay $\Lambda_Q\to
\Lambda_{Q'}$  are parameterized  in terms of six invariant form factors:
\begin{eqnarray}
  \label{eq:llff}
&&  \langle \Lambda_{Q'}(v')|V^\mu|\Lambda_Q(v)\rangle= \bar
  u_{\Lambda_{Q'}}(v')\Bigl[F_1(w)\gamma^\mu\cr
&&+F_2(w)v^\mu+F_3(w)v'^\mu\Bigl]
u_{\Lambda_Q}(v),\cr
&& \langle \Lambda_{Q'}(v')|A^\mu|\Lambda_Q(v)\rangle= \bar
  u_{\Lambda_{Q'}}(v')\Bigl[G_1(w)\gamma^\mu\cr
&&+G_2(w)v^\mu+G_3(w)v'^\mu\Bigl]
\gamma_5 u_{\Lambda_Q}(v),\qquad 
\end{eqnarray}
where   $u_{\Lambda_{Q}}(v)$ and
$u_{\Lambda_{Q'}}(v')$ are Dirac spinors of the initial and final
baryon with four-velocities $v$ and $v'$, respectively;
$q=M_{\Lambda_{Q'}}v'-M_{\Lambda_Q}v$,  and 
$w=v\cdot v'={(M_{\Lambda_Q}^2+M_{\Lambda_{Q'}}^2-q^2)}/
{(2M_{\Lambda_Q}M_{\Lambda_{Q'}})}.$

\begin{figure}
  \centering
  \includegraphics[height=6cm,angle=-90]{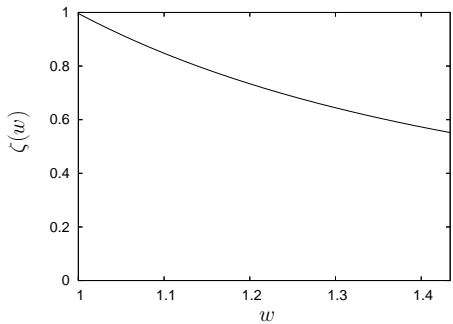}
  \caption{The Isgur-Wise function $\zeta(w)$ for 
the $\Lambda_b\to \Lambda_c e\nu$ semileptonic decay.} 
  \label{fig:xilbc}
\end{figure}

\begin{figure}
  \centering
  \includegraphics[height=6cm,angle=-90]{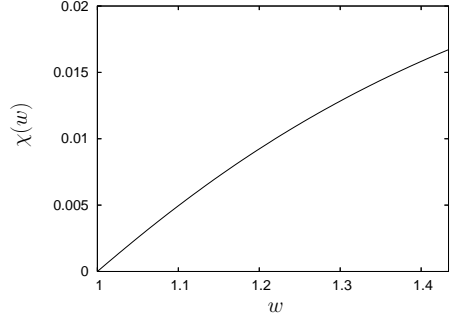}
  \caption{The subleading function $\chi(w)$ for
the $\Lambda_b\to
    \Lambda_c e\nu$ semileptonic decay.} 
  \label{fig:etalbc}
\end{figure}

In the heavy quark limit $m_Q\to\infty$ ($Q=b,c$) the form factors
(\ref{eq:llff}) 
can be expressed through the single Isgur-Wise function $\zeta(w)$\cite{iw} 
\begin{eqnarray}
  \label{eq:ffhl}
\!\!  F_1(w)&=&G_1(w)=\zeta(w),\cr
\!\!F_2(w)&=&F_3(w)=G_2(w)=G_3(w)=0.
\end{eqnarray}
At subleading order of the heavy quark expansion the form factors
are given by \cite{fn}
\begin{eqnarray}
  \label{eq:ffso}
  F_1(w)&=& \zeta(w)+\left(\frac{\bar\Lambda}{2m_Q}+
\frac{\bar\Lambda}{2m_{Q'}}\right)\Bigl[2\chi(w)\cr
&&+ \zeta(w)\Bigr], \qquad \bar \Lambda=M_{\Lambda_{Q}}-m_Q,\cr
 G_1(w)&=& \zeta(w)+\left(\frac{\bar\Lambda}{2m_Q}
+\frac{\bar\Lambda}{2m_{Q'}}\right)\Bigl[2\chi(w)\cr
&&+\frac{w-1}{w+1} \zeta(w)\Bigr],\cr
F_2(w)&=&G_2(w)=-\frac{\bar\Lambda}{2m_{Q'}}\frac{2}{w+1}\zeta(w),\cr
F_3(w)&=&-G_3(w)=-\frac{\bar\Lambda}{2m_{Q}}\frac{2}{w+1}\zeta(w),
\end{eqnarray}
where the leading order Isgur-Wise function of heavy baryons
\begin{eqnarray}
 \label{eq:iwf}
 &&\!\!\zeta(w)=\lim_{m_Q\to\infty}\int\frac{d^3p}{(2\pi)^3}\cr
&&\!\!\Psi_{\Lambda_{Q'}}\!\!
\left({\bf p}+2\epsilon_d(p)\sqrt{\frac{w-1}{w+1}}\ \bf e_\Delta\right)
\Psi_{\Lambda_{Q}}({\bf p}),\qquad
\end{eqnarray}
and the subleading  function
\begin{eqnarray}
  \label{eq:eta}
 &&\!\!  \chi(w)=-\frac{w-1}{w+1}\lim_{m_Q\to\infty}
\int\frac{d^3p}{(2\pi)^3}\frac{\bar \Lambda-\epsilon_d(p)}{2\bar \Lambda}\cr
&&\!\!\Psi_{\Lambda_{Q'}}\!\!
\left({\bf p}+2\epsilon_d(p)\sqrt{\frac{w-1}{w+1}}\ \bf e_\Delta \right)
\Psi_{\Lambda_{Q}}({\bf p}),\qquad
\end{eqnarray}
here $\bf e_\Delta={\bf \Delta}/\sqrt{{\bf \Delta}^2}$ is the unit vector in
the direction of   
${\bf \Delta}=M_{\Lambda_{Q'}}{\bf v'}-M_{\Lambda_{Q}}{\bf v}$,
$\epsilon_d(p)=\sqrt{{\bf p}^2+M_d^2}$. 
These functions, calculated with 
model wave functions for $\Lambda_b$ and $\Lambda_c$ baryons, are
plotted in Figs.~\ref{fig:xilbc},~\ref{fig:etalbc}. The function
$\chi(w)$ is very small in the whole accessible kinematic range, since
it is roughly proportional to the ratio of the heavy baryon binding energy
to the baryon mass.

Near the zero recoil point of the final baryon $w=1$ these
functions can be approximated by 
\begin{eqnarray}
  \label{eq:exp}
  \zeta(w)&=&1-\rho_{\zeta}^2(w-1)+c_{\zeta}(w-1)^2+\cdots,\cr
\chi(w)&=&\rho_\chi^2(w-1)+c_\chi(w-1)^2+\cdots,
\end{eqnarray}
where $\rho_{\zeta}^2=-[d\zeta(w)/dw]_{w=1}$ is the slope and
$2c_{\zeta}=[d^2\zeta(w)/d^2w]_{w=1}$ is the curvature of the Isgur-Wise
functions. Their values are given in Table~\ref{tab:sl}.

\begin{table}
  \tbl{Parameters of the Isgur-Wise functions for the
    $\Lambda_b\to\Lambda_c e\nu$ and $\Xi_b\to\Xi_c e\nu$ decays.
  \label{tab:sl}}
{\begin{tabular}{@{}cccccc@{}}
\toprule
Decay&$\bar\Lambda$ (GeV) &$\rho_{\zeta}^2$&$c_{\zeta}$&
$\rho_\chi^2$& $c_\chi$ \\
\colrule
$\Lambda_b\to\Lambda_c e\nu$& 0.764& 1.70& 2.39 & 0.053
&0.029\\
$\Xi_b\to\Xi_c e\nu$&0.970&2.27& 3.87&0.045&0.036\\
\botrule
\end{tabular}}
\end{table}

Our prediction for the branching ratio of the semileptonic decay
$\Lambda_b\to \Lambda_c e\nu$ for   $|V_{cb}|=0.041$ and
$\tau_{\Lambda_b}=1.23\times 10^{-12}$s \cite{pdg} 
$$Br^{\rm theor}(\Lambda_b\to \Lambda_c l\nu)=6.9\%$$
is in agreement with available experimental data\cite{delphi,cdf}
\begin{equation*}\label{dcdf}
Br(\Lambda_b\to \Lambda_c l\nu)=\left\{
\begin{array}{l} \left(5.0^{+1.1}_{-0.8}{}^{+1.6}_{-1.2}\right)\% 
 \cr
\left(8.1\pm  1.2^{+1.1}_{-1.6}\pm 4.3\right)\% 
\end{array}\right. \end{equation*}
and the PDG branching ratio \cite{pdg}
\begin{equation*}\label{dpdg}
Br(\Lambda_b\to \Lambda_c l\nu+{\rm
  anything})=(9.1\pm2.1)\%.
\end{equation*}

For the $\Xi_b\to\Xi_c e\nu$ decay we predict:
$$ Br(\Xi_b\to\Xi_c e\nu)=7.4\%.$$

The semileptonic decays of heavy baryons with the axial vector diquark
can be considered in the similar way. 
In the heavy quark limit $m_Q\to\infty$ the decay matrix element
is reduced to \cite{iw,bb}  
\begin{eqnarray}
  \label{eq:mhql}
   &&\!\!\langle \Omega^{(*)}_{Q'}(v')|\bar h_{v'}^{(Q')}\Gamma  
h_{v}^{(Q)}|\Omega_Q(v)\rangle= 
\bar B^{\Omega^{(*)}_{Q'}}_\mu(v')\Gamma\cr
&&\!\!
B^{\Omega_{Q}}_\nu(v)[-g^{\mu\nu}\zeta_1(w)+v^\mu v'{}^\nu \zeta_2(w)],
\end{eqnarray}
where 
\begin{eqnarray*}
  \label{eq:bb}
 &&
 B^{\Omega_{Q}}_\mu(v)=\frac1{\sqrt3}(\gamma_\mu+v_\mu)\gamma_5u_{\Omega_Q}(v),\cr
&& B^{\Omega^{*}_{Q}}_\mu(v)=u_{\Omega^*_Q,\mu}(v)
\end{eqnarray*}
and $u_{\Omega^*_{Q},\mu}$ is the Rarita-Schwinger spinor for
the $\Omega^*_{Q}$.

\begin{figure}
  \centering
  \includegraphics[height=6cm,angle=-90]{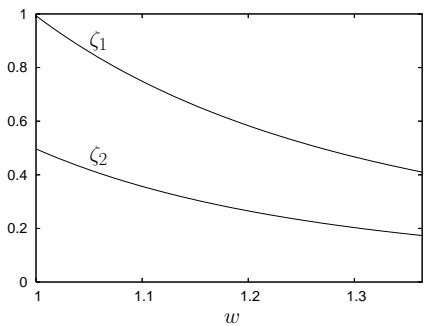}
  \caption{The Isgur-Wise functions $\zeta_1(w)$ and $\zeta_2(w)$ for 
the $\Omega_b\to
    \Omega^{(*)}_c e\nu$ semileptonic decay.} 
  \label{fig:zetao}
\end{figure}

The Isgur-Wise functions $\zeta_1(w)$ and $\zeta_2(w)$  are given by 
\begin{eqnarray}
  \label{eq:zeta}
&&\!\!  \zeta_1(w)=\lim_{m_Q\to\infty}
\int\frac{d^3p}{(2\pi)^3}\\
&&\!\!\Psi_{\Omega_{Q'}}\!\!
\left({\bf p}+2\epsilon_d(p)\sqrt{\frac{w-1}{w+1}}\ \bf e_\Delta\right)
\Psi_{\Omega_{Q}}({\bf p}),\cr
&&\!\!\zeta_2(w)=\frac1{w+1} \zeta_1(w), \label{eq:zeta2}
\end{eqnarray}
where $\bf e_\Delta={\bf \Delta}/\sqrt{{\bf \Delta}^2}$ is the unit
vector in the direction of  
${\bf \Delta}=M_{\Omega_{Q'}}{\bf v'}-M_{\Omega_{Q}}{\bf v}$. The
relation (\ref{eq:zeta2}) follows from the relativistic spin
rotation  of the spectator axial vector diquark. 
The Isgur-Wise functions are plotted in Fig.~\ref{fig:zetao}.

Near the zero recoil  point  $w=1$ the Isgur-Wise functions can 
again be approximated by 
\begin{equation}
  \label{eq:expzeta}
  \zeta_{i}(w)=\zeta_i(1)-\rho_{\zeta_i}^2(w-1)+c_{\zeta_i}(w-1)^2+\cdots,
\end{equation}
where $\zeta_1(1)=1$ and $\zeta_2(1)=1/2$; 
$\rho_{\zeta_i}^2=-[d\zeta_i(w)/dw]_{w=1}$ is the slope and
$2c_{\zeta_i}=[d^2\zeta_i/d^2w]_{w=1}$ is the curvature of the Isgur-Wise
functions. Their values are given in Table~\ref{tab:slzeta}.

\begin{table}
  \tbl{Parameters of the Isgur-Wise functions for the
    $\Sigma_b\to\Sigma^{(*)}_c e\nu$,  $\Xi'_b\to\Xi'{}^{(*)}_c e\nu$
    and $\Omega_b\to\Omega^{(*)}_c e\nu$ decays.
  \label{tab:slzeta}}
{\begin{tabular}{@{}cccccc@{}}
\toprule
Decay&$\bar\Lambda$ (GeV)   &$\rho_{\zeta_1}^2$&$c_{\zeta_1}$
&$\rho_{\zeta_2}^2$& $c_{\zeta_2}$ \\
\colrule
$\Sigma_b\to\Sigma^{(*)}_c e\nu$& 0.942& 2.17& 3.62& 1.34
&2.44\\
$\Xi'_b\to\Xi'{}^{(*)}_c e\nu$&1.082&2.61& 4.93&
1.55&3.19\\
$\Omega_b\to\Omega^{(*)}_c e\nu$&1.208&2.99& 6.21&
1.74&3.91\\
\botrule
\end{tabular}}
\end{table}

\begin{table*}
  \tbl{Comparison of different theoretical predictions for
    semileptonic decay rates $\Gamma$ (in $ 10^{10}{\rm s}^{-1}$) of
    bottom baryons. 
  \label{tab:tdr}}
{\begin{tabular}{@{}cccccccccc@{}}
\toprule
Decay&this work &[\refcite{singl}]& [\refcite{ct}]& [\refcite{kkp}] &[\refcite{ilkk}]
&[\refcite{iklr}]& [\refcite{cs}]&[\refcite{ahn}] &[\refcite{hjkl}]\\
\colrule
$\Lambda_b\to\Lambda_c e\nu$& 5.64& 5.9& 5.1& 5.14& 5.39& 6.09&
$5.08\pm 1.3$& 5.82 &$5.4\pm 0.4$\\
$\Xi_b\to\Xi_c e\nu$&5.29&7.2 & 5.3 & 5.21 & 5.27& 6.42& $5.68\pm 1.5$
& 4.98 &\\
$\Sigma_b\to\Sigma_c e\nu$& 1.44& 4.3 & &  & 2.23&1.65& & &\\
$\Xi'_b\to\Xi'_c e\nu$&1.34& & & & & & & &\\
$\Omega_b\to\Omega_c e\nu$& 1.29&5.4 & 2.3&1.52 &1.87&1.81& & &\\
$\Sigma_b\to\Sigma_c^* e\nu$& 3.23& & & &4.56 &3.75& & &\\
$\Xi'_b\to\Xi_c^* e\nu$ &3.09&  & & & & & & &\\
$\Omega_b\to\Omega_c^* e\nu$& 3.03& & & 3.41& 4.01&4.13 & & & \\
\botrule
\end{tabular}}
\end{table*}

The theoretical values for the decay rates of heavy
baryons containing scalar and axial vector diquarks and the comparison
of our model predictions with other theoretical 
calculations 
are given in
Table~\ref{tab:tdr}. In nonrelativistic quark models \cite{singl,ct,kkp}
form factors of the heavy baryon decays are evaluated at the single
kinematic point of zero recoil and then different form factor
parameterizations (pole, dipole) are used for decay rate calculations.   
The relativistic three-quark model \cite{ilkk}, Bethe-Salpeter
model \cite{iklr} and light-front constituent quark model \cite{cs}
assume Gaussian wave functions for heavy baryons. 
The authors of the recent nonrelativistic quark model \cite{ahn} use
for the form factor evaluations 
the set of variational wave functions, obtained from baryon spectra
calculations without employing the quark-diquark
approximation. Finally, Ref.~\refcite{hjkl} presents the recent QCD sum rule
prediction. Calculations of Refs.~\refcite{kkp,ilkk,iklr} are done in
the heavy quark limit only, while the rest include first order $1/m_Q$
corrections for the decays of $\Lambda$-type baryons. From
Table~\ref{tab:tdr} we see that all theoretical 
models give close predictions for the semileptonic decays of heavy
baryons with scalar diquark ($\Lambda_b\to\Lambda_c e\nu$ and
$\Xi_b\to\Xi_c e\nu$), which are consistent with the available
experimental data  for the
$\Lambda_b\to\Lambda_c e\nu$ semileptonic decay. Thus one can conclude
that the 
precise measurement of the semileptonic $\Lambda_b\to\Lambda_c e\nu$
decay rate will allow an accurate determination of the CKM matrix
element $V_{cb}$ with small theoretical uncertainties.

All predictions for heavy baryon decays with the axial vector diquark
listed in Table~\ref{tab:tdr} were obtained in the heavy quark
limit. Here the differences between predictions  are larger. The
nonrelativistic quark model \cite{singl} gives for these decay rates
values more than two times larger than other estimates. Our model
values for these decay rates are the lowest ones.
Among the relativistic quark models
the closest to our predictions is given in
\cite{iklr}. Unfortunately, it will be difficult to measure such
decays experimentally. Only  $\Omega_b$ (which has not been
observed yet) will decay predominantly weakly and thus 
has sizable semileptonic branching fractions, since a
scalar $ss$ diquark is forbidden by the
Pauli principle. All other baryons with the axial vector diquark will
decay predominantly strongly or electromagnetically and thus their
weak branching ratios will be very small. 

In summary, in this work we calculated the semileptonic decay rates
of heavy baryons in the framework of the relativistic quark
model. The baryon wave
functions were obtained previously in the process of the heavy baryon
mass spectrum calculations \cite{efghbm}. The  diquark size
is taken into account by calculating the diquark-gluon 
form factor.
This work was supported in part by the {\it Deutsche
Forschungsgemeinschaft} under contract Eb 139/2-3 and by the {\it Russian
Foundation for Basic Research} under Grant No.05-02-16243.

\end{document}